\documentclass[twocolumn]{aastex62}

\usepackage{xcolor}
\usepackage{yfonts}
\usepackage{amsmath}
\usepackage{graphicx}

\newcommand{\isois}{IS$\odot$IS }

\received{\today}

\shorttitle{Intermittent Heating}

\begin{document}

	\title{Observations of heating along intermittent structures in the inner heliosphere from PSP data}

\author[0000-0001-8358-0482]{R. A. Qudsi}
\affiliation{Department of Physics and Astronomy, University of Delaware, Newark, DE 19716, USA}
	
\author[0000-0002-2229-5618]{B.~A. Maruca}
\affiliation{Department of Physics and Astronomy, Bartol Research Institute, University of Delaware, Newark, DE 19716, USA}

\author[0000-0001-7224-6024]{W.~H. Matthaeus}
\affiliation{Department of Physics and Astronomy, Bartol Research Institute, University of Delaware, Newark, DE 19716, USA}

\author[0000-0003-0602-8381]{T.~N. Parashar}
\affiliation{Department of Physics and Astronomy, University of Delaware, Newark, DE 19716, USA}	

\author[0000-0002-6962-0959]{Riddhi Bandyopadhyay}
\affiliation{Department of Physics and Astronomy, University of Delaware, Newark, DE 19716, USA}

\author[0000-0002-7174-6948]{R. Chhiber}
\affiliation{Department of Physics and Astronomy, University of Delaware, Newark, DE 19716, USA}

\author[0000-0001-8478-5797]{A. Chasapis}
\affiliation{Laboratory for Atmospheric and Space Physics, University of Colorado Boulder, Boulder, CO 80303, USA}

\author[0000-0002-5317-988X]{Melvyn L. Goldstein}
\affiliation{NASA Goddard Space Flight Center, Greenbelt, MD 20771, USA}
\affiliation{University of Maryland Baltimore County, Baltimore, MD 21250, USA}

\author[0000-0002-1989-3596]{S.~D. Bale}
\affiliation{Space Sciences Laboratory, University of California, Berkeley, CA 94720-7450, USA}
\affiliation{Physics Department, University of California, Berkeley, CA 94720-7300, USA}
\affiliation{The Blackett Laboratory, Imperial College London, London, SW7 2AZ, UK}

\author{J.~W. Bonnell}
\affiliation{Space Sciences Laboratory, University of California, Berkeley, CA 94720-7450, USA}

\author{T. Dudok de Wit}
\affiliation{LPC2E, CNRS and University of Orl\'eans, Orl\'eans, France}

\author{K. Goetz}
\affiliation{School of Physics and Astronomy, University of Minnesota, Minneapolis, MN 55455, USA}

\author{P.~R. Harvey}
\affiliation{Space Sciences Laboratory, University of California, Berkeley, CA 94720-7450, USA}

\author{R.~J. MacDowall}
\affiliation{Code 695, NASA Goddard Space Flight Center, Greenbelt, MD 20771, USA}

\author{D. Malaspina}
\affiliation{Laboratory for Atmospheric and Space Physics, University of Colorado Boulder, Boulder, CO 80303, USA}

\author{M. Pulupa}
\affiliation{Space Sciences Laboratory, University of California, Berkeley, CA 94720-7450, USA}

\author{J.~C. Kasper}
\affiliation{Climate and Space Sciences and Engineering, University of Michigan, Ann Arbor, MI 48109, USA}
\affiliation{Smithsonian Astrophysical Observatory, Cambridge, MA 02138 USA}

\author[0000-0001-6095-2490]{K.~E. Korreck}
\affiliation{Smithsonian Astrophysical Observatory, Cambridge, MA 02138 USA}

\author[0000-0002-3520-4041]{A.~W. Case}
\affiliation{Smithsonian Astrophysical Observatory, Cambridge, MA 02138 USA}

\author[0000-0002-7728-0085]{M. Stevens}
\affiliation{Smithsonian Astrophysical Observatory, Cambridge, MA 02138 USA}

\author[0000-0002-7287-5098]{P. Whittlesey}
\affiliation{Space Sciences Laboratory, University of California, Berkeley, CA 94720-7450, USA}

\author{D. Larson}
\affiliation{Space Sciences Laboratory, University of California, Berkeley, CA 94720-7450, USA}

\author{R. Livi}
\affiliation{Space Sciences Laboratory, University of California, Berkeley, CA 94720-7450, USA}

\author{M. Velli}
\affiliation{Department of Earth, Planetary, and Space Sciences, University of California, Los Angeles, CA 90095, USA}

\author{N. Raouafi}
\affiliation{Johns Hopkins University Applied Physics Laboratory, Laurel, MD, USA}

		
		

	\begin{abstract}

        The solar wind proton temperature at 1-au has been found to be correlated with small-scale intermittent magnetic structures, i.e., regions with enhanced temperature are associated with coherent structures such as current sheets. Using Parker Solar Probe data from the first encounter, we study this association using measurements of radial proton temperature, employing the Partial Variance of Increments (PVI) technique to identify intermittent magnetic structures. We observe that the probability density functions of high-PVI events have higher median temperatures than those with lower PVI,  The regions in space where PVI peaks were also locations that had enhanced temperatures when compared with similar regions suggesting a heating mechanism in the young solar wind that is associated with intermittency developed by a nonlinear turbulent cascade.n the immediate vicinity.

	\end{abstract}


	\section{Introduction} \label{sec:intro}

        Solar wind is a stream of charged particles emanating from Sun originating in the corona \citep{Parker1960, Parker1963}. It is highly magnetized collisionless plasma streaming at supersonic speed and is primarily composed of ionized hydrogen (i.e., protons) \citep{Marsch1982, Kasper2012}.
        
        Despite decades of observation, the exact process that originally heats and accelerates solar-wind plasma remains unknown, but several candidates have been proposed. Turbulence cascade transfers energy from large to small scales, which can ultimate lead to dissipation and heating \citep{Velli1989, Velli1993, Matthaeus1999, Dmitruk2002, Cranmer2005, Cranmer2007, Cranmer2012, Cranmer2014, Verdini2007, Verdini2009, Verdini2009a, Chandran2009, Perez2013, Lionello2014}. Current sheets, generated by cascading vortices, can also lead to localized heating \citep{Parashar2009, Osman2011, Osman2012, Osman2012a, Gingell2015}. Wave particle interactions -- including, e.g., microinstabilities, Landau damping, and ion-cyclotron resonance -- can likewise result in heating and other dramatic changes to the particles' phase-space distribution \citep{Gary1993, Sahraoui2010, Klein2015}.
        
        
        In this study, we focused on coherent structures: features in the plasma that are persistent through time, concentrated in space, or both \citep{Greco2018}. Such structures can be produced by turbulent cascade \citep{Osman2012a} and are also associated with current sheets \citep{Yordanova2016}. \citet{Osman2011, Osman2012a} analyzed in-situ observations and of near-Earth solar wind and found clear indications that coherent structures correlate with local enhancements in temperature.
        
        In this study, we revisit the techniques of \citet{Osman2011, Osman2012a}, and, by applying them to observations from Parker Solar Probe (PSP), explore the relationship between plasma structures and heating in nascent solar-wind plasma. Section~\ref{sec:bgd} provides the background on such structures and introduces the reader to the physics of technique employed in data analysis which is described in section \ref{sec:data}. In section~\ref{sec:diss} we present the results and discuss its implication. Section~\ref{sec:conc} summarizes the results along with a conclusion and potential future works.

    \section{Background} \label{sec:bgd}

        The solar wind at 1-au exhibits localized structures that have been studied since the pioneering work of \citet{Burlaga1968, Hudson1970, Tsurutani1979} and others, and more recently by \citet{Ness2001, Neugebauer2006, ErdHoS2008}. Several studies have found evidence that plasma turbulence generates these structures dynamically \citep{Matthaeus1986, Veltri1999, Osman2013}. The structures are inhomogeneous and highly intermittent \citep{Osman2011, Osman2013, Greco2008} .
        
        Some recent studies, both observational and numerical, have shown that these structures are correlated with the regions of enhanced temperature in the plasma \citep{Osman2012, Osman2011, Greco2012} and understanding the mechanisms by which the turbulence heats the plasma may also help solve the coronal heating problem \citep{Osman2012a}. This is a particularly attractive scenario especially given the ubiquity of the localized structures. Study performed on data from PIC simulation by \citet{Wu2013} shows that the correlation between enhanced temperature and coherent structures exists for sub ion inertial length ($d_i$). Further evidence of this is provided by \citet{TenBarge2013} for Gyrokinetic simulation, \citet{Parashar2009, Wan2012, Karimabadi2013, Wan2015} for PIC and \citet{Servidio2012, Servidio2015} for Vlasov simulations respectively. Work done by \citet{Chasapis2015} and \citet{Yordanova2016} on  from Cluster and MMS data show similar results from observation vantage point.
        
        In this study, we investigate these discontinuities in the magnetic field and explore their association with local enhancements in ion temperature. One method for identifying a discontinuity in a time series of magnetic-field (or any other) data is Partial Variance of Increments (PVI), which is both a powerful and reliable tool for identifying and locating such regions. Following \citet{Greco2008} we define normalized PVI as:
		\begin{equation} \label{eqn:pvi}
		    \mathcal{I}(t, \Delta t) = \frac{|\Delta \mathbf{B}(t, \Delta t)|}{\sqrt{\langle |\Delta \mathbf{B}(t, \Delta t)|^2 \rangle}}
		\end{equation}
        where, $\Delta \mathbf{B}(t, \Delta t) = \mathbf{B}(t+\Delta t) - \mathbf{B}(t)$, is the vector increment in magnetic field at any given time $t$ and a time lag of $\Delta t$. For studying local structures induced by turbulence, $\Delta t$ is typically chosen to be, assuming the validity of Taylor's hypothesis \citep{Taylor1938}, of the order of $d_i$. $\langle ... \rangle$ is the ensemble average over a period of time, and $\mathcal{I}$ is the normalized PVI. \citet{Osman2012a} showed that PVI values greater than 2.4 imply the existence of strong non-Gaussian coherent structures. Although they constitute only a small fraction of total data set their contribution to the total internal energy per unit volume is high. This emphasizes the importance of using the PVI technique for such studies. We also note that an analogous examination of the association of PVI events with energetic particles was carried out at 1-au, \citep{Tessein2016}. A corresponding study of the first PSP orbit using \isois data is reported by Bandyopadhyay et al (this volume).

	\section{Data Selection and Methodology} \label{sec:data}
	
	    We analyzed data from PSP's first encounter with the Sun (October 31 to November 11, 2018). The FIELDS fluxgate magnetometers provided measurements of the local magnetic field at a rate of 64 samples/NYseconds, where 1 NYsecond is defined as 0.837 seconds \citep{Bale2016}. Radial proton temperature/thermal speed data was obtained from the Solar Probe Cup (SPC), part of the Solar Wind Electron, Alpha and Proton (SWEAP) suite \citep{Kasper2016}. The average speed of solar wind during the first encounter was around 350 km/s for the most part and crossed 500km/s only on the last day of the encounter. Thus, using Taylor’s Hypothesis, 1 NYs corresponds to a length scale of 300 km.
	    
	    Since SPC only measures radial temperature, and proton temperature is expected to be significantly anisotropic we needed to ensure that the temperature we were measuring was indeed parallel temperature. Thus, we only considered data points where magnetic field was mostly radial. Any interval where the angle between  $B_r \mathbf{\hat{r}}$ and $\mathbf{B}$ was more than 30 degrees was not considered. This ensured that the temperature measured by SPC was indeed the parallel temperature. For the calculation of PVI according to Equation~\ref{eqn:pvi}, we used 64 NYHz data, with a lag of 1 NYs, which is the native cadence of SPC \citep{Kasper2016}. The ensemble averaging was done over 8 hours, which is several times the estimated correlation time. In this study we used the correlation time computed in \citet{Parashar2019psp:prep}. However there are there are few subtleties associated with this calculation, and \citet{Smith2001, Isaacs2015, KrishnaJagarlamudi2019, Bandyopadhyay2019psp:prep}  offer more insights and discussion on this topic along with potential issues in such determination. We also carried out the analysis for various different averaging times (from 1 to 12 hours) and it was observed to have minimal affect on the outcome. PVI time series was then resampled to ion cadence of 1NYHz in the way such that for each interval of 1 NYs, the maximum value of PVI in that interval was chosen.
	    
	    In this study we focused on the second half of the encounter, immediately after PSP was at its perihelion. The second half of the encounter has very different properties compared to the first half. A greater number of energetic particles were observed \citep{McComas:sub}, the solar wind speed was higher \citep{Kasper2019:sub} , and there were many more switchbacks \citep{Bale2019:sub}. \citet{Bandyopadhyay2019psp:prep} observed enhanced local energy transfer, which points towards a more turbulent period in general, and thus a suitable environment for PVI study.

	\section{Results and Discussion} \label{sec:diss}
		\begin{figure}
			\begin{center}
				\includegraphics[width=0.5\textwidth]{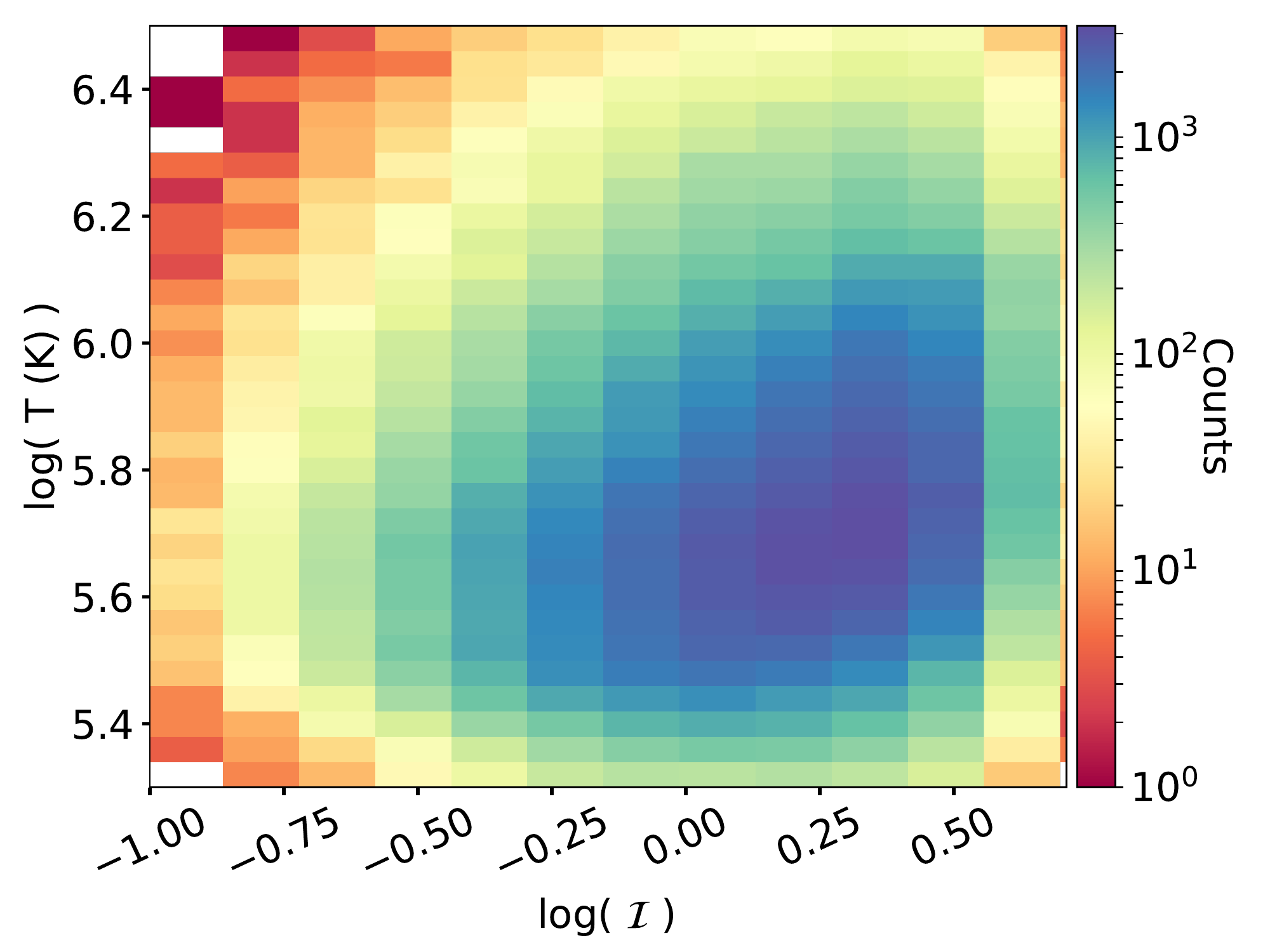}
				\caption{Joint histogram of radial proton-temperature and PVI for the second half of first encounter on a log-log scale. There is an upward trend between PVI and temperature as the blue region in the plot tilts upwards showing an increase of temperature as PVI increases.}
				\label{fig:pvi_hst}
			\end{center}
		\end{figure}

		\begin{figure}
			\begin{center}
				\includegraphics[width=0.5\textwidth]{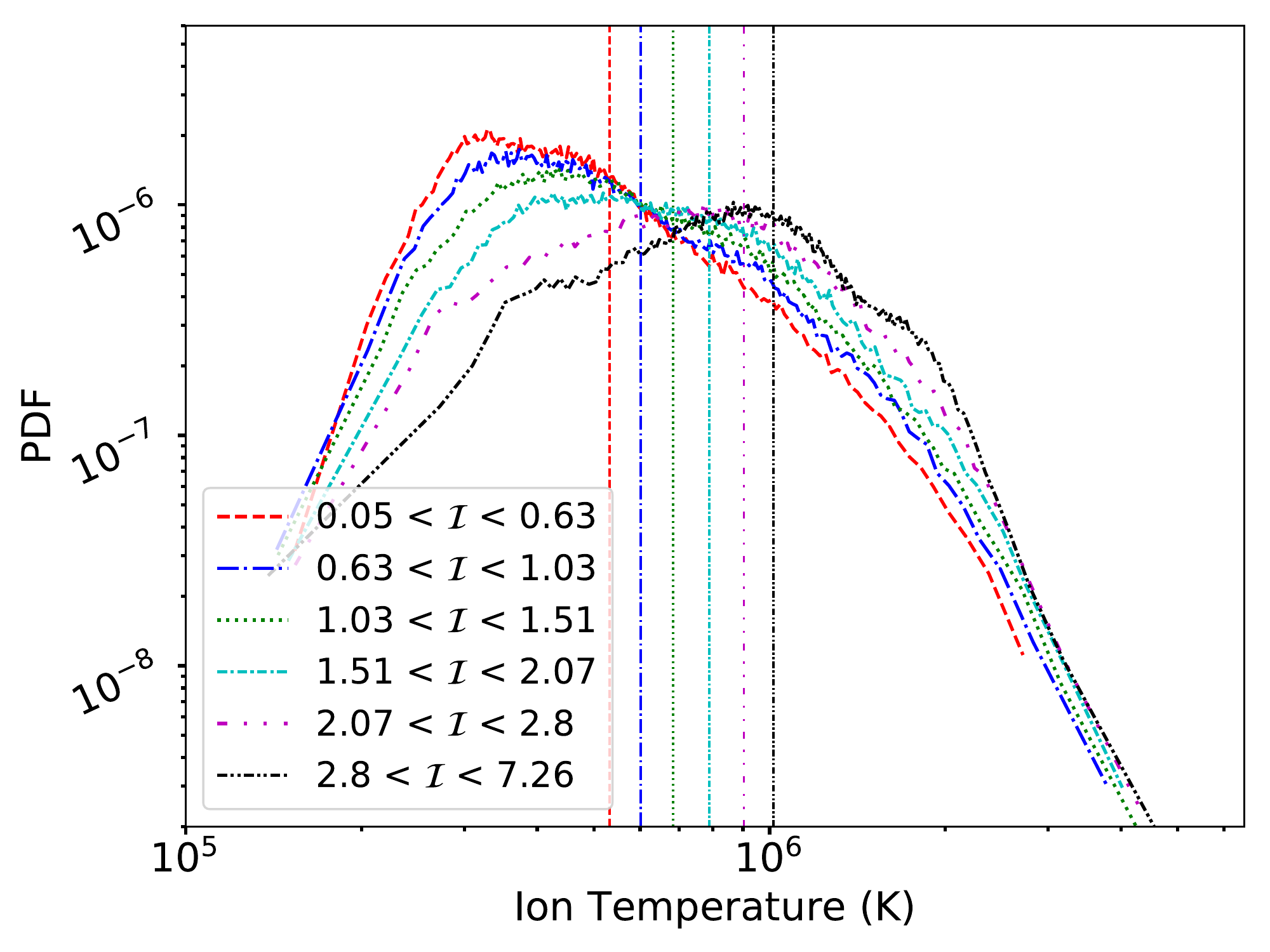}
				\caption{PDFs for the radial proton-temperature for the second half of first encounter. Each PDF corresponds to a different PVI range such that each PVI bin has equal number of data points. The probability density increases with increase in temperature for high PVI whereas it decreases for low PVI PDF. Vertical lines show the median temperature for each of the PDF plot.}
				\label{fig:pvi_pdf}
			\end{center}
		\end{figure}

	    Figure~\ref{fig:pvi_hst} shows the joint histogram of radial proton-temperature and PVI for the first encounter. Increasing PVI color contours have an upwards trend, as we see temperature distribution showing a positive slope with increase in value of PVI. The positive correlation between temperature and PVI suggest some kind of heating in the regions with high PVI. We then conditionally sampled radial proton temperature. Conditionally sampled means that we arrange the data by increasing value of PVI and then divide all the data points in 6 bins such that each bin has equal number of points. We then calculate the temperature distribution within each bin which is shown in Figure~\ref{fig:pvi_pdf}.

	    As PVI increases, the probability density increases for the higher temperature and decreases for the lower temperature which is opposite of what we see at the low temperatures where probability density is highest for the lowest PVI. Median temperature, shown by vertical lines in Figure~\ref{fig:pvi_pdf}, for each of the distribution increases implying presence of stronger and stronger heating as we go to higher and more extreme values of PVI. For PVI $<$ 1, median value of the temperature is $5.32 \times 10^5$ K whereas for PVI $>$ 6, the median temperature increases to $1.01 \times 10^6$ K. \citet{Osman2011} observed similar increase in average temperature in their study of solar wind at 1-au. This is consistent with heating occurring in the regions with small scale coherent structure in MHD turbulence.

	    In order to further demonstrate this relationship, we looked at the temperature at the point of high PVI event and its immediate surrounding in space using the methodology described by \citet{Osman2012a}. We compute the mean value of temperature at the point of the PVI event and for points near the PVI events separated from it by up to one correlation length. Formally, these averages may be expressed as:
        \begin{equation} \label{eqn:tts}
		    \widetilde{T}_p( \Delta t, \theta_1, \theta_2) = \langle T_p(t_{\tiny\mathcal{I}}+\Delta t)| \theta_1 \leq \mathcal{I}(t_\mathcal{I}) < \theta_2 \rangle
	    \end{equation}
        where $\widetilde{T}_p$ is the conditionally average temperature for all the events, $\Delta t$ is the time difference relative to the position of PVI events, $t_\mathcal{I}$ is the time of PVI events between the threshold $\theta_1$ and $\theta_2$

		\begin{figure}
			\begin{center}
				\includegraphics[width=0.5\textwidth]{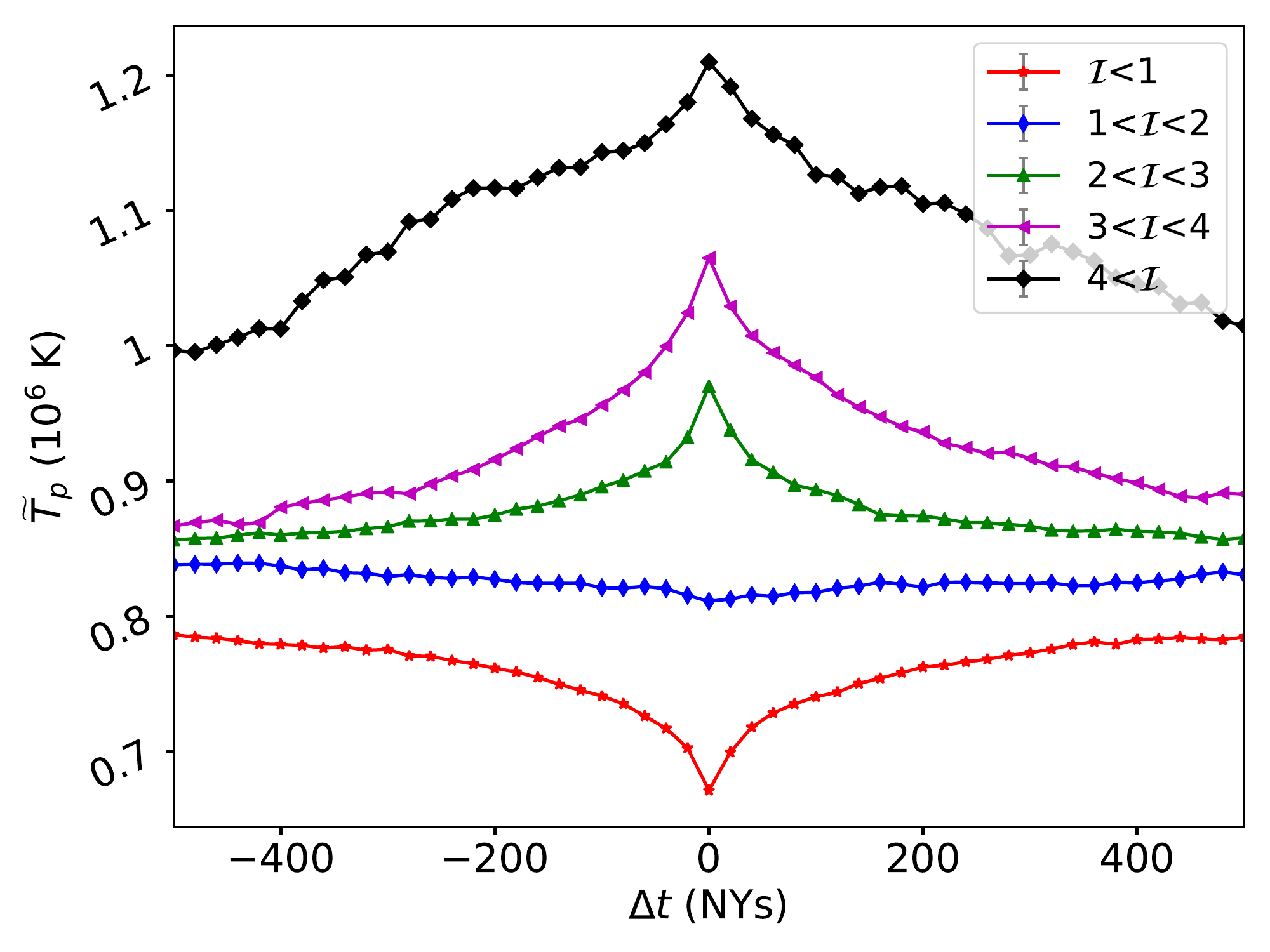}
				\caption{Figure shows conditional average temperature for different PVI thresholds at the point of a a PVI event. $\widetilde{T}_p$ peaks at the instant of PVI event and continues to have elevated temperature in its vicinity within the correlation time scale. Red curve, corresponding to lowest PVI shows a dip suggesting no heating when the magnetic field is very smooth.}
				\label{fig:tem_pvi_lag}
			\end{center}
		\end{figure}	

        Figure~\ref{fig:tem_pvi_lag} shows the plot of $\widetilde{T}_p$ for various thresholds for the second half of the encounter. Not only do we observe enhanced temperature at the point of high PVI events, suggesting localized heating at those points, we also see that $\widetilde{T}_p$ for a higher PVI event is consistently higher than nearby points separated by up to a correlation length. This implies that the points nearby an identified PVI event have an elevated average temperature, continuously approaching the elevated average temperature found at the PVI event itself. Some of this effect may be due to clustering of PVI events (see \citet{Chhiber2019psp:prep}). Another point worth noting in Figure~\ref{fig:tem_pvi_lag} is the valley in the temperature profile for small PVI. This is the region where background magnetic field is smooth and it appears that in such regions, the temperature is lower than the temperature of plasma in its immediate surrounding , which is concurrent with the fact that in those places there is no turbulence heating. \citet{Osman2012a} found similar result in their study at 1AU. However, in our study we find a significant dip compared to the dip reported in \citet{Osman2012a}, $\sim 10^5$ K compared to $\sim 2\times 10^3$ K.

	\section{Conclusion} \label{sec:conc}

        In this study, we used in-situ observations from PSP's first encounter with the Sun to explore the association of proton heating with coherent magnetic structures in the young solar wind. We identified enhancements of PVI \citep{Greco2008} as indicating the presence of such a structure \citep{Osman2011, Osman2012a}.  We observed that the joint histogram of PVI and proton radial temperature shows positive trend as shown in Figure~\ref{fig:pvi_hst}. We also observed that the PDF of data, as shown in Figure~\ref{fig:pvi_pdf} with higher PVI has higher mean temperature compared to those with lower PVI. This strongly supports the theory that the solar wind in those regions are heated by coherent structures which are generated by plasma turbulence.
        
        The present results demonstrate both the shifting of the PDF of  temperature towards higher  values with increasing PVI condition, (in Figure~\ref{fig:pvi_pdf})  and the spatial/temporal localization of the temperature enhancement near PVI events (in Figure~\ref{fig:tem_pvi_lag}).  Both of these are fully consistent with findings in the two papers that examine these effects ( \citet{Osman2011, Osman2012}, respectively) .  It is interesting that these effects are present clearly in the  PSP first orbit where turbulence is presumably younger and possibly less well developed than it is at 1 AU. It is possible that the temperature differential between low and high PVI is somewhat less in the PSP data than in the ACE data at 1 AU (Osman 2011), but additional samples by PSP will be needed to draw any firm conclusion of this type.

        In order to further demonstrate this association we looked at the conditionally average temperatures at the point of a high PVI event and in its immediate surrounding up to 1 correlation length. We observed that not only the point of event has the highest temperature, its vicinity shows enhanced temperature compared to lower PVI events. The local maxima of these temperature profiles are most prominent from higher PVI events suggesting stronger heating. The plateau region of each thresholds are distinct, and for higher threshold they maintain a high value suggesting clustering of PVI events around a large discontinuity. For very smooth magnetic field we see a dip in the average temperature at that point. \citet{Osman2012a} found similar behaviour in their study of solar wind at 1AU, though neither the heating nor the dip in temperature for small PVI that they reported in their study was as high as what we observed in our study. This suggests that either coherent structures are more efficient in heating the plasma near the Sun compared to 1-au or we have a lot more such structures as we move closer to the Sun. Since these coherent structures are generated by plasma turbulence, these observations suggest that non-linear turbulence cascade play a crucial role in heating the nascent solar wind. Given the ubiquitous nature of such structures, this process can help explain the coronal heating.
        
        A significant limitation of this study was unavailability of temperature-anisotropy data.  The temperature measures we used were not the scalar temperature but rather the radial temperature, for which reason we limited our observations to period of nearly-radial magnetic field (see Section~\ref{sec:data}).  Once ion temperature anisotropy data are available, we will revisit this work to explore both scalar and anisotropic heating.  Theoretical studies have found that turbulent cascade can generate strong temperature anisotropy near coherent structures \citep{Parashar2016}.
        
        We also hope to further explore Figure~\ref{fig:tem_pvi_lag}.  Careful inspection of this plot reveals a very slight asymmetry in the shape of the temperature profile before and after the PVI event. The phenomenon was also noted in 1-au solar wind by \citep{Osman2012a}. The cause and significance of this asymmetry remain unclear, but it may it suggests a connection between local heating and large-scale processes such as heat flux.

	\section*{Acknowledgment}

        This work was supported as a part of the PSP mission under contract NNN06AA01C. This research was partially supported by the Parker Solar Probe Plus project through Princeton \isois subcontract SUB0000165.

\end{document}